\documentclass{iopart}
\usepackage{graphicx}
\usepackage{iopams}

\begin{document}

\title[Thermomodulation studies of heat transport]{Thermomodulation studies of heat transport in heavy-doped n-GaAs by using the
temperature dependence of metal-semiconductor contact resistance}
\author{A.Ya. Shul'man, N.A. Mordovets, and I.N. Kotel'nikov}

\address{Institute of Radio Engineering and Electronics of the RAS\\Moscow
125009, Russia. E-mail: ash@cplire.ru}

\begin{abstract}
The heat transport in heavy-doped n-GaAs has been investigated at temperatures
$T=300$ K and $77$ K using the irradiation of the metal-semiconductor contact
by modulated CO$_{2}$-laser radiation. It is shown this approach giving an
opportunity to determine the thermo-diffusion coefficient $\chi$ and Seebeck
coefficient $S_{T}$ without direct measurements of the temperature gradient.
It was also found out that the thermalization length of hot electrons exceeds
in of order of magnitude the assessment which can be done based on the
reference data for GaAs. To elucidate the origin of the observed phenomenon
the measurements were conducted out with Schottky contacts made on the thin
doped GaAs layer epitaxially grown on the semi-insulating GaAs substrate. In
this case the degenerate electron gas occupies only insignificant part of the
heat-conducting medium. In addition, the injection of hot electrons into the
semiconductor by current pulses through the Schottky barrier was used to clear
up whether there is a dependence of the effect on the method of the electron
heating. The nonequilibrium of LO-phonons and the change in the
electron-phonon collisional integral due to the non-equilibrium pair
correlations of the electrons are suggested as a possible explanation.
\footnote{Presented in part on 13th International Conference
''Nonequilibrium Carrier Dynamics in Semiconductors'' (HCIS-13), Modena,
Italy, 2003 and 6th Russian Semiconductor Physics Conference, S-Peterburg, 2003}
\end{abstract}

%Uncomment for PACS numbers title message
%

%Uncomment for Submitted to journal title message
%\submitto{\JPA}
%
%

%Comment out if separate title page not required
%\maketitle
%
%

\section{Experiment and data handling}

\begin{figure}
[th]
\begin{center}
\includegraphics[
height=7.5146cm,
width=8.211cm
]%
{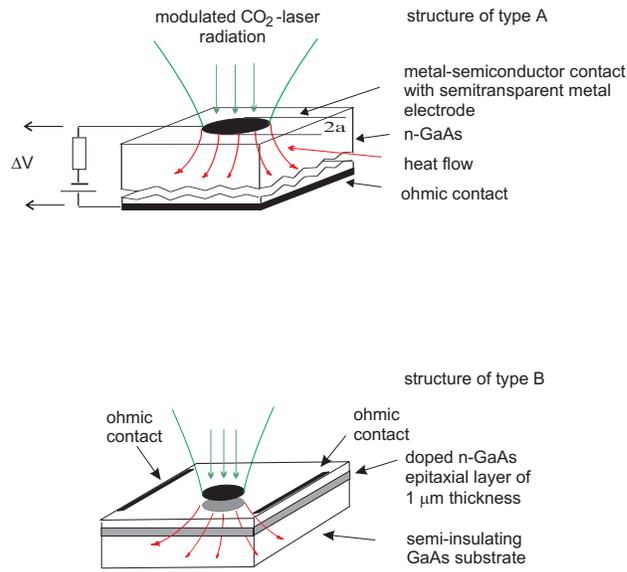}%
\caption{A draft of measurement scheme and different types of samples'
layout}%
\label{Fig1}%
\end{center}
\end{figure}

The frequency dependence of thermal response of the degenerate electron gas in
GaAs (the Fermi energy $\sim80-130$ meV) has been studied at
spatially-nonuniform heating. The surplus energy of the electron gas was
produced by free carriers absorption of chopped $10$-$\mu m$ laser radiation
(see Fig.\ref{Fig1}).%

%TCIMACRO{\FRAME{fthFU}{8.211cm}{7.5146cm}{0pt}{\Qcb{A draft of measurement
%scheme and different types of samples' layout}}{\Qlb{Fig1}}{fig_emf_c.eps}%
%{\special{ language "Scientific Word";  type "GRAPHIC";
%maintain-aspect-ratio TRUE;  display "ICON";  valid_file "F";  width 8.211cm;
%height 7.5146cm;  depth 0pt;  original-width 4.9294in;
%original-height 4.5083in;  cropleft "0";  croptop "1";  cropright "1";
%cropbottom "0";  filename 'Fig_emf_c.eps';file-properties "XNPEU";}}}%
%BeginExpansion
%EndExpansion

\begin{figure}
[th]
\begin{center}
\includegraphics[
height=6.8227cm,
width=8.2417cm
]%
{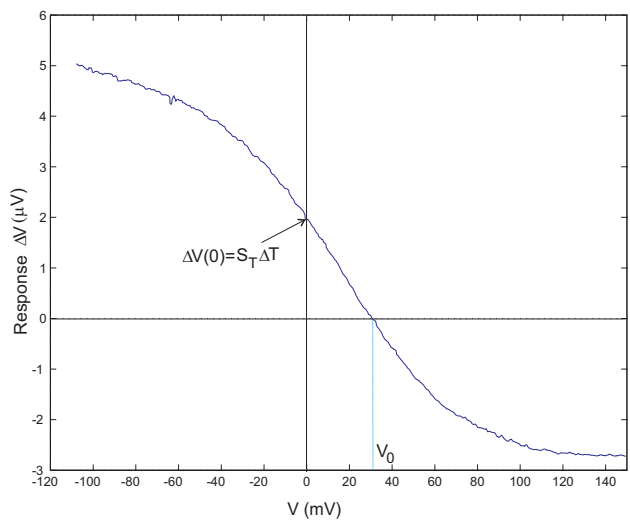}%
\caption{Bias dependence of the thermoresponse to CO$_{2}$-laser radiation.
The dependence of the response voltage $\Delta V$ on the chopper frequency $F$
was the subject of the investigation. The bias voltage $V_{0}$ where the
response changes in its sign can be used to determine the Seebeck coefficient
$S_{T}$ according to the Eq.(\ref{Seeb-main}). Experimental parameters:
electron density $N=7.0\cdot10^{18}\,$cm$^{-3}$, incident radiation power
$P=250\,$mW, modulation frequency $F=570\,$Hz, environment temperature
$T=77\,$K.}%
\label{Fig2}%
\end{center}
\end{figure}

The opposite side of the sample was held at the equilibrium temperature. The
change in the electron temperature $T$\ close to the metal-semiconductor
interface led to change in the metal-semiconductor tunnel junction resistance
and to the nascent thermoelectromotive voltage. The plot in Fig.\ref{Fig2}
demonstrates the dependence of the response $\Delta V$ on the voltage bias $V$
measured by phase-sensitive detector. At $V=0$ the response is the sole
thermoelectromotive signal. At $V\neq0\,\ $the response is the sum of the
thermo-e.m.f. and the change in the $V$ in the case of non-zero current $I$
through the junction due to the temperature dependence of the junction
resistance $R$.%

%TCIMACRO{\FRAME{fthFU}{8.2417cm}{6.8227cm}{0pt}{\Qcb{Bias dependence of the
%thermoresponse to CO$_{2}$-laser radiation. The dependence of the response
%voltage $\Delta V$ on the chopper frequency $F$ was the subject of the
%investigation. The bias voltage $V_{0}$ where the response changes in its sign
%can be used to determine the Seebeck coefficient $S_{T}$ according to the
%Eq.(\ref{Seeb-main}). Experimental parameters: electron density $N=7.0\cdot
%10^{18}\,$cm$^{-3}$, incident radiation power $P=250\,$mW, modulation
%frequency $F=570\,$Hz, environment temperature $T=77\,$K.}}{\Qlb{Fig2}%
%}{fig1ppc.eps}{\special{ language "Scientific Word";  type "GRAPHIC";
%maintain-aspect-ratio TRUE;  display "ICON";  valid_file "F";
%width 8.2417cm;  height 6.8227cm;  depth 0pt;  original-width 5.8496in;
%original-height 4.8326in;  cropleft "0";  croptop "1";  cropright "1";
%cropbottom "0";  filename 'Fig1ppc.eps';file-properties "XNPEU";}}}%
%BeginExpansion
%EndExpansion

\subsection{Bias dependence of thermoresponse and determination of Seebeck coefficient}

Such an experimental setup allows us to suggest a simple way of measurement of
the Seebeck coefficient $S_{T}$ for the degenerate electron gas in the
heavy-doped semiconductor when the standard methods give a large spread of
values (see e.g. \cite{Nasl-73}). A simple circuit small-signal analysis with
the current-driven junction gives the condition for the bias $V_{0}$ where
$\Delta V=0$:
\begin{equation}
\Delta V\equiv\left[  V_{0}\left(  \frac{\partial\ln R}{\partial T}\right)
_{I}-S_{T}\right]  \Delta T=0.\label{Seeb-main}%
\end{equation}
Here $\Delta T$\ is the temperature difference averaged over the metal contact
area. Evidently, it is possible to determine $S_{T}$ from Eq.(\ref{Seeb-main})
without $\Delta T$ measuring. All other quantities involved in this
relation\ characterize\ the junction itself and can be easily measured. As an
example, the value of $V_{0}=32$ mV presented in the Fig.\ref{Fig1} being
combined with the value of $d\ln R/dT\simeq-10^{-3}$\ K$^{-1}$ gives
$S_{T}\simeq-30$ $\mu$V/K in good agreement with the value expected for the
indicated electron density and temperature. In fact, the derivation  of
Eq.(\ref{Seeb-main}) is based on the potentiality of the temperature- and
Coulomb fields and on the constancy of the Coulomb potential along all metal
contacts including ohmic ones \cite{Sh-2be}. This allows to prove the
Eq.(\ref{Seeb-main}) is independent of geometry of  the current and heat flows
in the samples as long as the junction resistance is much greater than the
bulk resistance of the substrate. Therefore, both type of the structures shown
in Fig.\ref{Fig1} have been suitable for that sort of measurements.

%TCIMACRO{\FRAME{fhFU}{8.1099cm}{8.9447cm}{0pt}{\Qcb{Comparative plot of
%electron density dependence of Seebeck coefficient in n-GaAs at $T=300$ K.
%Selected by color data are obtained with the technique suggested in present
%work.}}{\Qlb{Fig3}}{nasledov_engm.eps}{\special{ language "Scientific Word";
%type "GRAPHIC";  maintain-aspect-ratio TRUE;  display "ICON";
%valid_file "F";  width 8.1099cm;  height 8.9447cm;  depth 0pt;
%original-width 6.3304in;  original-height 6.9851in;  cropleft "0";
%croptop "1.0003";  cropright "1.0001";  cropbottom "0";
%filename 'Nasledov_engM.eps';file-properties "XNPEU";}}}%
%BeginExpansion
\begin{figure}
[bh]
\begin{center}
\includegraphics[
trim=0.000000in 0.000000in -0.000633in -0.002096in,
height=8.9447cm,
width=8.1099cm
]%
{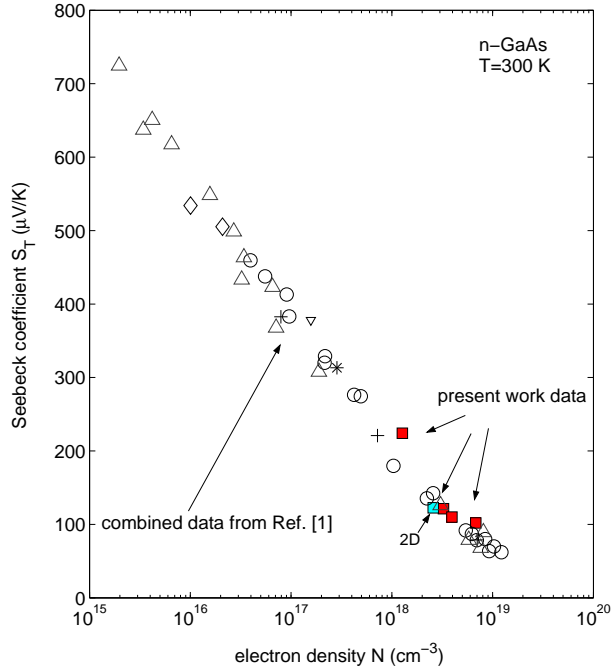}%
\caption{Comparative plot of electron density dependence of Seebeck
coefficient in n-GaAs at $T=300$ K. Selected by color data are obtained with
the technique suggested in present work.}%
\label{Fig3}%
\end{center}
\end{figure}
%EndExpansion

The Seebeck coefficients for several structures with various values of
electron density measured by this technique are presented in the Table and
compared with known data in Fig.\ref{Fig3}.

Table. Parameters of structures and respective Seebeck
coefficients\nolinebreak \newline
\begin{tabular}
[c]{|c|c|c|c|c|c|c|c|}\hline
{\footnotesize preparation} & {\footnotesize structure} & $N${\footnotesize ,}%
& $R_{300K}${\footnotesize ,} & $ d\ln R/dT,$ &
$V_{0},$ & $S_{T},${\footnotesize  } & $\Delta
T_{\symbol{126}}$, K$%
%TCIMACRO{\UNICODE{0xb0}}%
%BeginExpansion
{{}^\circ}%
%EndExpansion
${\footnotesize  }\\
{\footnotesize technology} & {\footnotesize type} & {\footnotesize 10}$^{18}%
${\footnotesize cm}$^{-3}$ & {\footnotesize KOhm} & {\footnotesize 10}$^{-3}/$%
{\footnotesize K}$%
%TCIMACRO{\UNICODE{0xb0}}%
%BeginExpansion
{{}^\circ}%
%EndExpansion
$ & {\footnotesize mV} & $\mu ${\footnotesize V/K}$%
%TCIMACRO{\UNICODE{0xb0}}%
%BeginExpansion
{{}^\circ}%
%EndExpansion
$ & {\footnotesize (F=600 Hz)}\\\hline
MOCVD & B & 1.27 & 124.0 & -6.49 & 34.4 & -223 & 0,03\\\hline
MBE & B & 3.25 & 0.945 & -17.8 & 6.8 & -121 & 0,07\\\hline
MBE & B & 3.95 & 5.590 & -11.1 & 9.93 & -110 & 0,06\\\hline
zone-melting & A & 6.80 & 0.004 & -7.64 & 13.4 & -102 & 0,11\\\hline
MBE & B & 2.60\footnotemark & 7.480 & -6.38 & 19.2 & -123 & 0,03\\\hline
\end{tabular}%
\footnotetext{Equivalent volume density of 2D electrons with
$N_{2D}=9\times10^{11}$ cm$^{-2}$ in $\delta$-doped layer of order of $3.5$ nm
width.}
\subsection{Modulation frequency dependence of thermoresponse and
determination of hot-electron thermalization length}

\begin{figure}
[th]
\begin{center}
\includegraphics[
height=11.2621cm,
width=7.1039cm
]%
{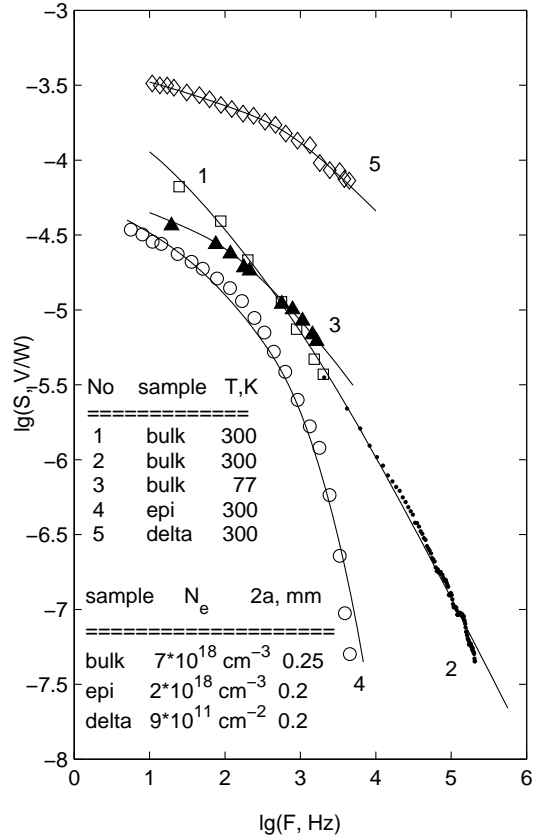}%
\caption{The dependence of the heating signals $S$ on the frequency of
CO$_{2}$-laser beam chopper for several tunnel Schottky-barrier junctions with
various electron densities $N_{e}$ and diameters of the metal contact $2a$
(curves 1, 3-5). Curve 2 was measured using Q-switched CO$_{2}$ laser and the
posterior Fourier-transform data processing. Markers denote experimental data.
Solid curves have been calculated from Eq.(\ref{Resp-base}) with parameters
described in the text.}%
\label{Fig4}%
\end{center}
\end{figure}

The measured frequency dependences of the thermoresponse for three junctions
at two ambient temperatures are shown in Fig.\ref{Fig4} by markers. The curves
1-3 were measured with the same junction made on the bulk-doped GaAs
substrate. The curves 4 and 5 were measured with the junctions made on the
epitaxially-doped side of the semi-insulating GaAs substrate. The thicknesses
of the substrates varied in the $0.3\div0.8$ mm range. The thickness of the
doped epi-layer of the sample for curve 4 was about $1$ $\mu$m. The junction
relating to the curve 5 was formed by surface-deposited metal film and 2D
electron gas in the $\delta$-doped channel placed at the $20$ nm distance from
the semiconductor surface.%

%TCIMACRO{\FRAME{fthFU}{7.1039cm}{11.2621cm}{0pt}{\Qcb{The dependence of the
%heating signals $S$ on the frequency of CO$_{2}$-laser beam chopper for
%several tunnel Schottky-barrier junctions with various electron densities
%$N$ and diameters of the metal contact $2a$ (curves 1, 3-5). Curve 2 was
%measured using Q-switched CO$_{2}$ laser and the posterior Fourier-transform
%data processing. Markers denote experimental data. Solid curves have been
%calculated from Eq.(\ref{Resp-base}) with parameters described in the text.}%
%}{\Qlb{Fig4}}{fig2m.eps}{\special{ language "Scientific Word";
%type "GRAPHIC";  maintain-aspect-ratio TRUE;  display "ICON";
%valid_file "F";  width 7.1039cm;  height 11.2621cm;  depth 0pt;
%original-width 3.4748in;  original-height 5.5279in;  cropleft "0";
%croptop "1";  cropright "1";  cropbottom "0";
%filename 'fig2m.eps';file-properties "XNPEU";}}}%
%BeginExpansion
%EndExpansion

It is noticeably that curves 4 and 5 show more slow-down behavior than the
curve 1 as a function of the modulation frequency $F$\ in the low-frequency
region up to $F\sim10^{3}$ Hz. In fact, this dependence is closer to
$1/F^{1/2}$ for the curve 4 and 5 and to $1/F$ for the curve 1. More
quantitative comparison of the measured curves with the heat transport theory
has been carried out on the base of the theoretical analysis presented in
\cite{Sh-97}.

For the sake of simplicity of the discussion let us consider the corresponding
expression for the response which takes into account all essential details of
the measurements but the finite thickness of the substrate:%
\begin{equation}
\Delta T(\Omega)=\frac{P(\Omega)(1-r)K}{\pi a\kappa}\int_{0}^{\infty}%
ds\frac{\exp(-s^{2}b^{2}/2)J_{1}(sa)J_{0}(sd)}{\sqrt{i\Omega/\chi+s^{2}%
}\left(  \sqrt{i\Omega/\chi+s^{2}}+K\right)  } \label{Resp-base}%
\end{equation}
Certainly, the real calculations have been conducted with the full more
complicated formula. Here $P(\Omega)$ is the power of the incident radiation
beam, $\Omega=2\pi F$, $r$ is the radiation reflection coefficient, $a$ is the
radius of the metal contact, $b$ is the dispersion of the Gaussian
distribution of the radiation power across the beam in the focal spot, $d$ is
the distance along the sample surface between the centre of the metal contact
and the focal spot,$\ \kappa$ is the semiconductor heat conductivity and $K$
is the free carrier absorption coefficient. In the case of heat transport in
homogeneous medium the quantity $k_{\Omega}=\sqrt{i\Omega/\chi}$ is the root
of dispersion equation. At the real $\Omega$ the $Re$ %
$1/k_{\Omega}$ plays the role of the wavelength and the $Im$ %
$1/k_{\Omega}$ is the attenuation length of the heat wave.

For the inhomogeneous heating under consideration the frequency dependence of
the right-hand side of Eq.(\ref{Resp-base}) at $d=0$ is determined an
interrelation between $\left|  k_{\Omega}\right|  $ and reminder of the scale
parameters $a,b,K$. The case $d\neq0$ must be considered separately.

The qualitative analysis of the expression (\ref{Resp-base}) as a function of
$\Omega$ with account for the actual limitation of the integration region by
the damping exponential- or oscillating Bessel function (or both) in the
integrand shows that at factual values of parameters $a$ and $b$ of order
$200-250$ $\mu$m and $1/K$ of order of $20-50$ $\mu$m the response must down
as $\sim1/\sqrt{F}$ in the range of $10^{2}$ Hz$<F<10^{4}$ Hz and transforms
to $\sim1/F$ law at higher frequencies. All these estimations are made at
$T=300$ K and reference value of the thermodiffusion coefficient $\chi
_{300}=0.26$ cm$^{2}$/s obtained from thermoconductivity $\kappa_{300}=0.46$
Wcm$^{-1}$K$^{-1}$ and heat capacity $C_{300}=1.75$ Jcm$^{-3}$K$^{-1}$of GaAs
given in \cite{Blak-82} and \cite{HandB-75}. Respective values for $T=77$ K
are $\kappa_{77}=3.2$ Wcm$^{-1}$K$^{-1},C_{77}=0.85$ Jcm$^{-3}$K$^{-1},$
$\chi_{77}=3.8$ cm$^{2}$/s .

The observed behavior of the curve 1-3 in Fig.\ref{Fig4} is evidently in
contradiction with the reasoning presented above and exact computations of
Eq.(\ref{Resp-base}) support this evidence. Since the experimental data is
represented in the form of amplitude of oscillating response, the thin
theoretical curves in Fig.\ref{Fig4}\ represents the absolute value of the
Eq.(\ref{Resp-base}). The fitted solid curves show results of calculations
with optimal values of the thermodiffusivity coefficient $\chi$ and the
reference values \cite{Nasl-73,Blak-82} of the radiation absorption
coefficient $K$. 1-2: $\chi=0.026$ cm$^{2}$/s, $K=450$ cm$^{-1}$, 3:
$\chi=0.38$ cm$^{2}$/s, $K=450$ cm$^{-1}$. The theoretical curves 4 and 5 were
calculated at the reference values of $\chi=0.26$ cm$^{2}$/s and $K=200$
cm$^{-1}$. These results prove that the response of the samples where
degenerate electron gas occupies only small part of the substrate volume can
be described with the standard value of the thermodiffusivity $\chi$.
Otherwise (curves 1-3), the values of $\chi$ were needed to decrease in 10
times. The very sharp drop of the curve 4 in the high-frequency tail is the
result of large displacement ($250\mu$m) between the centers of the focused
laser beam ($b=45\mu$m) and the metal electrode. Such a kind of measurements
allows to determine $\chi$ directly disregarding uncertainties of other
parameters involved as it will farther be explained.

By inspecting of Fig.\ref{Fig4} one can see that exact calculations of
Eq.(\ref{Resp-base}) with the indicated reference values of the parameters
$\chi$ and $K$ enable to describe the measured frequency dependence for the
junctions on the epitaxially-grown substrates only (curves 4 and 5). Just in
this case the heat transport passes mainly through the part of the
semiconductor substrate that is free of electrons. In the other case,
presented by curves 1-3, the used values of $\chi$ were reduced in 10 times to
reach an agreement with experimental data under all measurement conditions.
Almost the same results can be obtained with the reference values of $\chi$
and reduced in 10 times $K$. The meaning of this interchangeability will be
discussed later. It follows that just the electronic constituent is
responsible for the anomalous heat propagation in the semiconductor.

To check if the observed effect depends on the way of the electron heating,
the pulsed current heating at the reverse- and the forward polarities was
realized by specially designed switching electronic circuit. The circuit
enabled to measure the thermoresponse signal of order of $200$ nV in $50-100$
$\mu$s after the voltage pulse of $1$ V magnitude was supplied, in spite of
the large capacitance $\sim2-5$ nF of the investigated junctions. The
time-dependent pulsed response measured by the boxcar integrator was converted
then in the frequency domain using digital Fourier transformation.

It turn out that frequency dependences obtained at the reverse and forward
pulsed bias are different. At the forward bias the non-equilibrium electrons
were generated inside the metal film and the emitted phonons only were primary
mechanism for the heat transport into the semiconductor. In that case, the
calculated frequency dependence of the response manifested $\varpropto
1/F^{1/2}$ drop-down behavior up to several hundred kHz corresponding to the
near-surface heat generation and the normally expected lattice heat transport
into the semiconductor. At the reverse bias electrons were injected across the
Schottky barrier from the metal electrode into the semiconductor with kinetic
energy about $1$ eV above Fermi level and produced highly non-equilibrium
state of the electron gas due to intensive electron-electron collisions. That
is the Joule heat was directly scattered inside the electron gas of the
semiconductor. The respective frequency dependence shows the main sign of the
abnormal heat transport, e.g., the fall off more sharp then $1/F^{1/2}$
beginning from several hundred Hz.

In principle, the thermoresponse can fall off with the frequency more quickly
then $1/F$ even under the normal heat transport conditions as it is seen from
the experimental and calculated curves 4 in Fig.\ref{Fig4}. That case takes
place for $d\neq0$, i.e., when there is a distance between the centers of the
metal electrode and the focal spot. Surely, this behavior is contained in the
Eq.(\ref{Resp-base}) and can be explained on that ground. However, there is a
simpler way to understanding that gives us also some practically useful means.
The solution of the heat conductivity equation far away the excitation region
has the asymptotic form like $\exp(ik_{\Omega}d)/d$. Taking into account the
definition of $k_{\Omega}$ given above it is easy to obtain the dependence of
the response $S$ on the frequency $\Omega$ at $d$ constant in the form%
\begin{equation}
\ln S\varpropto-d\sqrt{\Omega/2\chi}.\label{Resp-far}%
\end{equation}

Accordingly to this expression there is the linear dependence of $\ln S$ on
$\Omega^{1/2}$. That dependence can be applied to the determination of $\chi$
without any reference to the junction (heat detector) characteristics and
irradiation conditions excluding the distance $d$. The application of the
claimed method to the curve 4 results in the value of $\chi\approx0.23$
cm$^{2}$/s in a good agreement with the adopted value $0.26$ cm$^{2}$/s.

\section{Discussion}

The observed anomalous frequency dependence of the thermal response can be
related either to a diminution of an effective thermal diffusivity in the
system of electrons + LO-phonons + acoustic phonons or to a real increasing of
the size of the region where the thermalization of the nonequilibrium electron
- LO-phonon subsystem takes place owing to the interaction with the thermal
bath of acoustic phonons. The investigated frequency band of the heating
modulation is too low-frequency for one to expect diminution of the thermal
diffusivity owing to the mechanism examined in \cite{Volz-01}. The change in
$\chi=\kappa/C$ may be attributed either change in $\kappa$ or in $C$ or in
both. Since the heat conductivity $\kappa$ appears in Eq.(\ref{Resp-base}) in
front of the integral besides $\chi$ we can use the absolute magnitude of the
response to distinguish the contributions of $\kappa$ and $C$ in the observed
change of thermodiffusivity $\chi$. The realization of this procedure shown
that the main role in the change of $\chi$ belong to the heat capacitance, if
this change does exist. The joint set of heat conductivity equations for the
electrons, LO-phonons, and acoustic phonons analyzed in \cite{Sh-97} allows
supposing that the first case might be considered as the result from rising
the contribution of the nonequilibrium electron - LO-phonon subsystem to the
heat capacity of the semiconductor.

The second case may be taken into consideration because the calculated curves
consist also with the measured data 1-3 for the bulk-doped sample in
Fig.\ref{Fig4} at thermodiffusivity $\chi_{300}=0.26$ cm$^{2}$/s and
$\chi_{77}=3.8$ cm$^{2}$/s (reference values) and reduced radiation absorption
coefficient $K=50$ cm$^{-1}$ instead of $K=500$ cm$^{-1}$. The value
$\chi=0.26$ cm$^{2}$/s has been confirmed by direct measurements of the
thermoresponse as a function of the intercenter distance $d$ according to
Eq.(\ref{Resp-far}). The observed phenomenon can then be interpreted as a
long-range tail effect in the spatial response of an inhomogeneous electron
gas corresponding to the known long-time tail in the time response of
spatially-homogeneous electron gas. As it is known, the latter is related with
the nonequilibrium corrections to the pair correlation function of electrons.
In \cite{Sh-02} the expression for the electron energy loss rate via
electron-phonon interaction was derived with explicitly found contribution of
the nonequilibrium pair correlations. However, the role of this correction in
the heat transport is unclear yet.

\ack
%\begin{acknowledgement}
%
%
We are thankful to V.B Sandomirsky, V.A. Sablikov and I.B. Levinson for
several discussions of the photothermospectroscopy and unconventional phonon
thermoconductivity while looking for the nonexotic explanation of the observed
effect. The partial financial support of Russian Foundation for Basic
Researches gratefully appreciates (grants No. 03-02-16728 and 03-02-16756.).
%\end{acknowledgement}
%
%

\end{document}